\documentclass[twocolumn,prl,showpacs,amsmath,amssymb]{revtex4}
\usepackage{graphicx}
\usepackage{dcolumn}
\usepackage{bm}
\usepackage{float}
\begin{document}
\title{On collisions driven negative energy waves and Weibel instability of a
relativistic electron beam in a quasi-neutral plasma}
\author{Anupam Karmakar$^1$}
\author{Naveen Kumar$^1$}
\author{Gennady Shvets$^2$}
\author{Oleg Polomarov$^2$}
\author{Alexander Pukhov$^1$}\email{pukhov@tp1.uni-duesseldorf.de}
\affiliation{$^1$Institut f\"ur Theoretische Physik I, 
Heinrich-Heine-Universit\"at, D\"usseldorf, 40225, Germany \\
$^2$Department of Physics and Institute for Fusion Studies, University of Texas
at Austin, One University Station, Austin, TX 78712, USA}

\begin{abstract}
A new quasi-neutral model describing the Weibel instability of a high-current
relativistic beam propagating through a resistive plasma is developed.
It treats beam electrons as kinetic particles, and ambient
plasma as a non-relativistic fluid. For a finite-temperature beam, a new class of
negative energy magneto-sound waves is identified, which can possess
negative energy. Their growth due to collisional dissipation in the
cold return current destabilizes the beam-plasma system even for high
beam temperatures. We perform detailed two- and three-dimensional particle-in-cell
(PIC) simulations of the thermal beam and collisional plasma. It is
shown that in three dimensions, the Weibel instability persists even
for physically collisionless background plasma. The anomalous
plasma resistivity is then caused  by the two-stream instability.
\end{abstract}
\maketitle

The fast ignition fusion (FI) is a promising route towards the laser driven
fusion. In the classic FI scheme \cite{tabak94}, a laser-generated relativistic
electron beam with a few MeV per electron energy must propagate through
overdense plasma to heat a hot spot in the core of a pre-compressed fusion fuel
target. The current carried by these MeV electrons inside the plasma is much
higher than the Alfv\'en current limit $I=(m c^3/e )\gamma= 17\gamma\, k$A,
where $m$ is the electron mass, $e$ is the electronic charge, and $\gamma$ is
the Lorentz factor of the beam. Transportation of this electron beam is not
possible unless it is compensated by a return plasma current. However, this
configuration is unstable and the current beam is subject to the Weibel and the
two-stream instabilities. The Weibel instability \cite{weibel59} is particularly
responsible for the generation of very strong static magnetic fields ($\sim
100$ MG) \cite{pukhov96}. It is one of the leading instabilities
under relativistic conditions and has been studied for a long time
\cite{pukhov97,deutsch05,jung05,silva02,
bret05,honda2000,honrubia05,honrubia06,califano06}.
Honrubia \emph{et al.} \cite{honrubia06} have performed three-dimensional
simulations of resistive beam filamentation corresponding to the full scale FI
configuration. Three-dimensional magnetic structures generated due to the Weibel
instability in a collisionless plasma have also been reported \cite{califano06}.
Recently, the evidence of Weibel-like dynamics and the resultant filamentation
of electron beams have been reported experimentally \cite{jung05}.
It was proposed in Ref.~\cite{silva02} that this instability could be suppressed
by the transverse beam temperature alone in a collisionless plasma. However,
the instability persists in the presence of collisions in return plasma current
no matter how high transverse beam temperature is. This regime of instability
was termed as resistive beam instability~\cite{molvig75}.

In this Letter, we develop a theoretical approach to the collisional Weibel
instability in the framework of the quasi-neutrality assumption. For a
finite-temperature beam, a new class of negative energy magneto-sound waves is 
identified. We derive conservation laws for the beam-plasma system and show that
the energy of this system is not positively definite. Rather, it contains a
negative term that allows for negative energy waves. Collisions in the
background plasma current excite unstable magneto-sound waves in the system,
which carry negative energy densities. Due to these waves the Weibel instability
is not suppressed even when the transverse beam temperature would be high enough
to stabilize collisionless plasma. 

We present results of detailed 2D and 3D particle-in-cell (PIC) simulations on the
relativistic electron beam transport in plasmas. The 2D geometry corresponds to
a plane transverse to the beam propagation direction. In this geometry the
Weibel instability is decoupled from the two-stream instability and we can study
the effects of temperature and collisions on the Weibel instability
systematically. The simulations results show that the Weibel instability cannot
be suppressed by thermal effects alone if collisions are present in the system.
We also make 3D PIC simulations of the Weibel instability. The simulation
results show that in the full 3D geometry, the Weibel instability cannot be
suppressed even in plasma free from binary collisions. We conjecture that the
effective collisions leading to an anomalous resistivity in the return current
are provided by the turbulence emerging from the electrostatic beam instability.

In the 2D simulations, we assume a very long electron beam propagating in the
$Z-$direction and there is no dependence on the coordinate $z$. The beam and
plasma densities are $n_b$ and $n_p$ respectively and the beam-plasma system is
quasi-neutral \emph{i.e.}~$n_b+n_p=n_0,\text{where}\,n_0$ is the background ion
density. Initially, the beam current is completely neutralized by the return
plasma current. However, as the Weibel instability develops the current
neutrality is destroyed due to the filamentation. The strongest magnetic field
is generated in the transverse plane ($x-y$ plane). This magnetic field
generates an axial component of the electric field $E_z$. The transverse
components of the electric field can be easily obtained from the force
equilibration $\bm E+\upsilon_{pz} \times \bm B_{\perp}/c=0$, where
$\upsilon_{pz}=\upsilon_{pz}\bm e_z$ is the return current velocity. To
summarize, these are the dominant electric and magnetic fields of the beam
plasma system: 

\begin{equation}
\bm B_{\perp}=-\bm e_z\times\nabla_{\perp} A_z,\,E_z=-\frac{1}{c}\frac{\partial
A_z}{\partial t},\,\bm E_{\perp}=-(\upsilon_{pz}/c)\nabla_{\perp} A_z,
\end{equation}

\noindent where $A_z$ is the $z$ component of the vector potential. The $B_z$
component is small, but can be approximated using $\partial_t
B_z=-\left(\nabla_{\perp}\upsilon_{pz}\times\nabla_{\perp}A_z\right)
\cdot \bm e_z$. The $A_z$ component is determined from the Ampere law 

\begin{equation}\label{fields}
\nabla_{\perp}^2 A_z=-\frac{4 \pi}{c}\left(J_{bz}+J_{pz}\right),
\end{equation}

\noindent where $J_{bz}$ and $J_{pz}$ are the current densities of the beam and
plasma respectively. We discard the displacement current to ensure the
quasi-neutrality. The axial equation of motion for the plasma flow and the
transverse equations of motion for beam electrons are 

\begin{equation}\label{pflow}
\frac{\partial \upsilon_{pz}}{\partial t}+\nu \upsilon_{pz}=\frac{e}{mc}\frac{\partial A_z}{\partial t},
\end{equation}

\begin{equation}\label{bflow}
\frac{d(\gamma_j\upsilon_{j\perp})}{d t}=-\frac{e(\upsilon_{jz}-\upsilon_{pz})}{mc}\nabla_{\perp}A_z,
\end{equation}

\noindent where $\nu$ is the collisional frequency of the ambient plasma, $m$
and $c$ are the electron mass and velocity of light in vacuum respectively, and
the subscript $j$ represents the $j^{th}$ beam electron. For collisionless
plasma, Eq.~\eqref{bflow} is written as 

\begin{equation}\label{bflow1}
\frac{d(\gamma_j\upsilon_{j\perp})}{d t}=-\frac{e\upsilon_{jz}}{mc}\nabla_{\perp}A_z+\frac{e^2}{2m^2c^2}\nabla_{\perp}A_z^{2}.
\end{equation}

\noindent The second term in the RHS of Eq.~\eqref{bflow1} is due to the extra
pinching of the electron beam by the transverse electric field $\bm E_{\perp}$.
We note here that $\bm E_{\perp}$ counters the magnetic expulsion of the ambient
plasma. At the same time it reinforces magnetic pinching of the beam. The
generalized momentum conservation in the $z-$direction gives

\begin{equation}\label{cons}
\gamma_j \upsilon_{jz}=\gamma_{j0}\upsilon_{jz0}+\frac{e}{mc}\left(A_z-A_{z0}\right).
\end{equation}

\noindent If there is no dissipation, then we may derive conservation laws for
the system. From Eqs.~\eqref{fields},~\eqref{bflow1} and \eqref{cons}, we have 

\begin{multline}\label{laws}
\sum_j \gamma_j mc^2-\sum_j \frac{m}{2}\left(\frac{e A_z}{mc}\right)^2+\int d^2x L_z \frac{|\bm \nabla A_z|^2}{8\pi}\\ +\int d^2x L_z \frac{n_0 m}{2}\left(\frac{e A_z}{m c}\right)^2=0,
\end{multline}

\noindent where $L_z$ is the system length in the $z$ direction. The first term in the above expression represents the total beam electron energy. The third and
fourth terms correspond to the total magnetic energy and the plasma kinetic
energy. The fourth term slightly overestimates the energy because the actual
plasma density $n_p=n_0-n_b$ is slightly smaller. The second term corrects
this overcount: the excess energy is subtracted from the electron beam energy.

\begin{figure}[htb]
\centering
\includegraphics[width=0.45\textwidth,height=0.55\textwidth]
{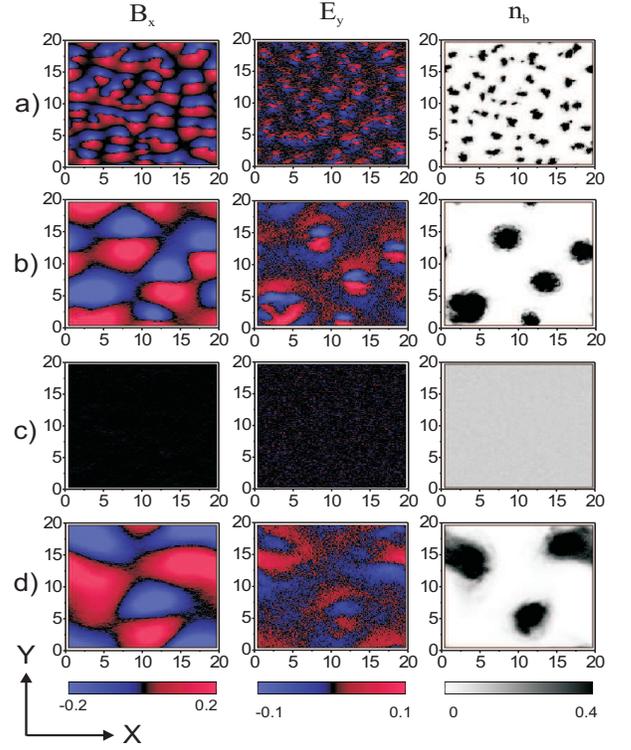}
\caption{(color online)~Snapshots of the the evolution of transverse
electromagnetic Weibel fields ( $E_x$ and $B_x$) and beam filament densities
($n_b$) during the nonlinear stage at a time $T=20(2\pi/\omega_{pe})$ for four
different simulation cases: (a) Cold electron beam in a collisionless background
plasma, (b) cold $e$-beam in a collisional background plasma and (c) hot
electron beam in a collisionless background plasma and (d) hot electron beam in
a collisional background plasma.}
\label{snapshots}
\end{figure}

The relativistic treatment of the instability could be somewhat cumbersome.
However, the essential physics can be learnt from the non-relativistic equation
of motion. For a warm electron beam the equation of motion reads as 

 \begin{equation}\label{eom}
 m\frac{d\bm \upsilon_{b\perp}}{d t}=-e\frac{\upsilon_{bz}-\upsilon_{pz}}{c}\nabla_{\perp}A_z-\frac{\nabla_{\perp}P}{n_b},
 \end{equation}

\noindent where $P$ is the beam pressure related to the beam emittance. On
linearizing Eq.~\eqref{eom} for small magnetic field perturbation, we have 

\begin{equation}\label{pertb}
\frac{d\delta\bm \upsilon_{b\perp}}{d t}=-c^2\beta_0\nabla_{\perp}\tilde{A_z}-3\frac{\upsilon_{th}^2}{n_{b}}\nabla_{\perp}\delta n_b,
\end{equation}

\noindent where $\tilde{A_z}=e
A_z/mc^2,\,\beta_0=(\upsilon_{bz}-\upsilon_{pz})/c\approx\upsilon_{bz}/c$,
and it was assumed that $\nabla_{\perp}\delta
P=3\upsilon_{th}^2\nabla_{\perp}\delta n_b$. Eq.~\eqref{pertb}
together with the continuity equation $\partial_t\delta
n_b=-n_{b}\left(\nabla_{\perp}\cdot\delta\upsilon_{b\perp}\right)$
yields 

\begin{equation}\label{sound}
\left(\frac{\partial^2}{\partial t^2}-c_s^2\nabla_{\perp}^2\right)\frac{\delta n_b}{n_{b}}=c^2\beta_0\nabla_{\perp}^2 \tilde{A_z},
\end{equation}

\noindent where $c_s^2=3\upsilon_{th}^2$ is the square of the beam's sound
speed. Eqs.~\eqref{fields},~\eqref{pflow}, and \eqref{sound} form a set of
equations to describe the sound like perturbation/filamentation of the electron
beam density. The simplest case of collisionless plasma ($\nu=0$) and long
wavelength perturbation ($\mid k_\perp\mid\ll\omega_{p}^2/c^2$, $\omega_p$ being
the ambient plasma frequency) gives

\begin{equation}\label{sound1}
\left(\frac{\partial^2}{\partial t^2}-\bar{c}_s^2\nabla_{\perp}^2\right)\frac{\delta n_b}{n_{b}}=0,
\end{equation}

\noindent where $\bar{c}_s^2=c_s^2-c^2\beta_0^2n_{b}/n_0$ is the modified sound
speed. One may note that the cold beam  ($c_s^2 < c^2\beta_0^2n_{b}/n_0$) is
unstable due to the Weibel instability whereas the warm beam is stable. Under
the warm beam approximation the dispersion for sound waves is given by
$\omega^2=\bar{c}_s^2 k_\perp^2$. These waves are stable and the Weibel
instability does not occur for sufficiently high transverse beam temperature and
low beam/plasma density ratios. With finite plasma resistivity taken into
account the dispersion relation for the sound waves reads as 

\begin{equation}\label{sound2}
\omega^2=c_s^2 k_{\perp}^2-\frac{\omega_{b}^2\beta_0^2 k_\perp^2}{(k_\perp^2+k_{pe}^2/(1+i\nu/\omega))}.
\end{equation}

\noindent where $\omega_b$ is the beam plasma frequency. For large scale
perturbations ($k_\perp^2 \ll k_{pe}^2$, $k_{pe}^{-1}=c/\omega_p$) and small
collision frequency, Eq.~\eqref{sound2} yields two damped modes and one growing
mode given by 

\begin{equation}
\omega\approx\pm\bar{c}_s k_\perp-i\nu\frac{c^2\beta_0^2n_{b}}{2\bar{c}_s^2n_0},
\end{equation}
\noindent and
\begin{equation}\label{nega}
\omega\approx i\nu\frac{c^2\beta_0^2n_{b}}{\bar{c}_s^2n_0}.
\end{equation}

\noindent Thus collisions drive negative energy waves
in the system, leading to the Weibel instability of a warm electron beam, which
would be stable in collisionless plasmas. This result has a profound impact on
the understanding of the Weibel instability in plasmas.  

To check our analytic theory, we carry out detailed 2D PIC simulations. The
relativistic electron beam propagates in the negative $\hat{z}$-direction with
the initial velocity ${\upsilon_{b,z}}$. The compensating return current  of the
ambient plasma electrons flows with the initial velocity ${\bf{\upsilon_p}}$.
The  plasma ions are immobile and have the density $n_0=n_b +n_{p}$.  The
simulation domain has the size $X\times
Y=\left(20\,\lambda_s\times20\,\lambda_{s}\right)$, where
$\lambda_{s}=c/\omega_{pe}$ is the plasma skin length. All simulations are
performed with 64 particles per cell and with a grid size of, $\delta x=\delta
y=0.125\,\lambda_s$. The density ratio between the beam and plasma electrons is
$n_{p}/n_b=9$, whereas the beam and background plasma electrons have velocities
$\upsilon_b=0.9\,c$ and $\upsilon_p=0.1\,c$.  The binary collisions are
simulated with a newly implemented collision module in the relativistic PIC code
Virtual Laser Plasma Laboratory  (VLPL)~\cite{vlpl99}. We record the evolution
of field energy for every component $F_i$ of the fields $\bf E$ and $\bf B$ as
$\int_S\left({e F_i} {m_e c\,\omega_{pe}}\right)^2 ~dx~dy$ where ${e
F_i}/{m_ec\omega_{p}}$ represents the relativistic field normalization. 

We use the electron beam initial temperature of $T_{ b}\sim 70$ keV and the
ambient plasma  collision frequency $\nu_{ei}/\omega_{p}=0.15$ for these
simulations.  

\begin{figure}[htb]
\includegraphics[width=0.45\textwidth,height=0.4\textwidth]
{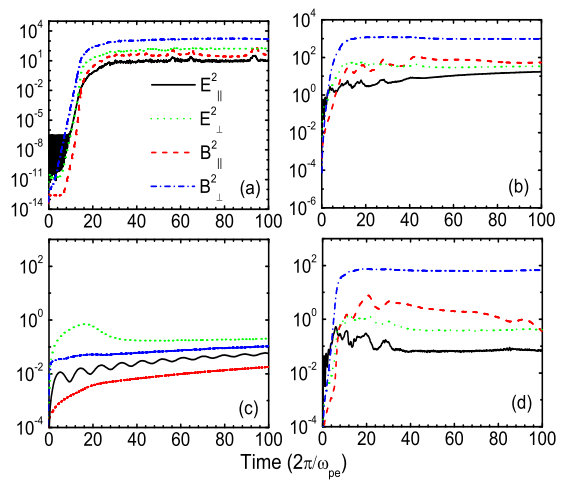}
\caption{(color online) Time evolution of the perpendicular and parallel Weibel
${\bf E}$ ${\bf B}$ field energies ($E^2_{\perp}$, $B^2_{\perp}$,
$E^2_{\parallel}$, $B^2_{\parallel}$) for four different simulation cases 
as described in the previous Fig.~\ref{snapshots}.}
\label{fig:fields}
\end{figure}

Fig.~\ref{snapshots} shows snapshots of the transverse $E$ and $B$ fields, and
the structure of the beam filaments at the time, $T=20(2\pi/\omega_{pe})$ for
four different cases: (a) cold electron beam and collisionless background
plasma, (b) cold electron beam and collisional plasma (c) hot electron beam and
collisionless plasma and (d) hot electron beam and collisional background
plasma. The beam density filamentation is shown in the last column in
each panel. In collisionless case (a), the filaments are small, comparable with
the background plasma electron skin depth. In the collisional case (b), the
filament size is bigger. This can be explained as a collisional diffusion of
plasma electrons across the self-generated magnetic fields.  In the third panel
of figure, simulation case (c), the electron beam is hot with the transverse
temperature $T_b=70$ keV, and the background plasma electrons are collisionless.
Here we see no filament formation. The temperature of the electron beam
stabilizes the  Weibel instability. Physically the thermal pressure of the
electron beam prevails over the magnetic pressure in this case. Hence, the
magnetic field pinching which actually drives the instability does not occur
resulting in the suppression of the Weibel instability. The last panel of the
figure depicts the filament formation in the simulation case (d), where electron
beam is hot and the plasma is collisional. Although the beam temperature is the
same as in the stable collisionless case (c), the background plasma collisions
revive back the instability. This is due to the collision induced generation of
negative energy waves as discussed earlier in the theoretical model.

\begin{figure}[htb]
\includegraphics[width=0.4\textwidth,height=0.3\textwidth]{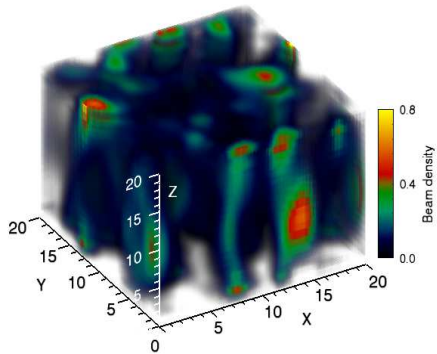}
\caption{(color online) Beam filaments from 3D PIC simulations. The ambient 
plasma is cold and collisionless,  while the electron beam is hot and 
collisionless. The other parameters are same as in Fig.~\ref{snapshots}(c).}
\label{3dfilaments}
\end{figure}

\noindent Fig.~\ref{fig:fields} shows the evolution of electric and magnetic
field energies in the four cases corresponding to the simulations in
Fig.~\ref{snapshots}. The energy axises in Fig.~\ref{fields} use logarithmic
scales. We see a stage of linear instability, where the field energies build 
up exponentially in time. It is followed by a nonlinear saturation. The linear
instability stage is present in the simulations (a), (b) and (d). The simulation
(c), where the electron beam had high temperature and the background plasma was
collisionless, shows no linear instability and no significant build up of the
magnetic field energy.

The linear growth rates calculated from the simulations results agree well with
the theoretical model. After the linear stage of the instability, an electrostatic regime of the filamentation instability characterized by the magnetic attraction of the filaments starts and the the field energies saturate.
Some small fluctuations around the saturated field energies can be seen. These
fluctuations occur due to the collective merging of the filaments as also
discussed in~\cite{honda2000}.

We also have done a number of 3D PIC simulations of Weibel instability,
varying the beam temperature and the plasma collision frequency. To
our surprise, we found no stabilization even in the collisionless case
for high beam temperatures. The corresponding simulation is shown in
Fig.~\ref{3dfilaments}. Although the electron beam in this simulation had the
high transverse temperature, and the plasma had no binary collisions,
we see a lot of filamentation due to the Weibel instability. We
explain this fact in terms of anomalous plasma collisionality. Indeed, there is
an oblique mode in the 3D geometry which couples the Weibel and the two-stream
instabilities~\cite{bret05}. The two-stream mode generated electrostatic
turbulence in the plasma. Stochastic fields associated with this turbulence
scatter the beam and plasma electrons and lead to an effective collisionality in
the return current. This anomalous effect revives back the Weibel instability.
It may be noted here that we have always taken the background plasma as cold.
The allowance of finite background plasma temperature decreases the growth rate
of the instability. The interplay of collisions in different regimes of beam and
background plasma temperatures can be found in Refs.\cite{deutsch05}.

In summary, we have developed a simplified model which identifies the
collisional Weibel instability as the instability of the unstable negative
energy mode driven by the collisions in the background plasma. An important
result of this study is that beam temperature does not  kill the
Weibel instability in the presence of collisions in beam plasma system. An
alternate explanation on the persistence of the Weibel instability in 3D
geometry is offered.  It is attributed to the anomalous collisionality of the
beam-plasma system due to the two-stream mode. We have also derived the
conservation laws for the beam plasma system which are useful for benchmarking
the numerical codes. Detailed 2D simulations on the Weibel instability of an
electron beam in two-dimensional transverse geometry have been performed, which
essentially confirm the theoretical prediction.

This work was supported by the DFG through TR-18 project and the US Department
of Energy grants DE-FG02-04ER41321 and DE-FG02-04ER54763.

\end{document}